\documentclass[useAMS,usenatbib]{mn2e}
\usepackage{longtable}
\usepackage{amssymb}
\usepackage{aas_macros}
\usepackage{natbib}
\usepackage{times}
\usepackage[dvips]{graphicx}

\title[Observed Cluster Concentration-Mass Relation]{The Observed
  Concentration-Mass Relation for Galaxy Clusters} 

\author[Comerford \& Natarajan]{Julia M. Comerford$^{1}$ and
  Priyamvada Natarajan$^{2,3}$\\
$^{1}$Astronomy Department, 601 Campbell Hall, University of
California, Berkeley, CA 94720-3411 \\
$^{2}$Department of Astronomy, Yale University, P.O. Box 208101, New
  Haven, CT 06520-8101 \\
$^{3}$Department of Physics, Yale University, P.O. Box 208120,
  New Haven, CT 06520-8120}

\begin{document}

\maketitle

\begin{abstract}
The properties of clusters of galaxies offer key insights into the
assembly process of structure in the universe. Numerical simulations of
cosmic structure formation in a hierarchical, dark matter dominated
universe suggest that galaxy cluster concentrations, which are a
measure of a halo's central density, decrease gradually with virial
mass. However, cluster observations have yet to confirm this
correlation. The slopes of the run of measured concentrations with
virial mass are often either steeper or flatter than predicted by
simulations. In this work, we present the most complete sample of
observed cluster concentrations and masses yet assembled, including
new measurements for 10 strong lensing clusters, thereby more than
doubling the existing number of strong lensing concentration
estimates. We fit a power law to the observed concentrations as a
function of virial mass, and find that the slope is consistent with
the slopes found in simulations, though our normalization factor is
higher. Observed lensing concentrations appear to be systematically
larger than X-ray concentrations, a more pronounced effect than found
in simulations.  We also find that at fixed mass, the bulk of observed
cluster concentrations are distributed log-normally, with the
exception of a few anomalously high concentration clusters. We examine
the physical processes likely responsible for the discrepancy between
lensing and X-ray concentrations, and for the anomalously high
concentrations in particular. The forthcoming Millennium simulation
results will offer the most comprehensive comparison set to our findings
of an observed concentration-mass power law relation.
\end{abstract}


\section{Introduction}
\label{intro}

Galaxy clusters are the most recent structures to assemble in a
hierarchical $\Lambda$CDM universe and therefore offer important clues
to the detailed understanding of the growth of structure in the
universe. The $\Lambda$CDM paradigm is well studied and
observationally supported on the largest scales by the cosmic
microwave background data, high-$z$ supernovae, and galaxy
surveys. Clusters of galaxies are a useful laboratory to further test
this paradigm, as their masses, abundances, and other properties such
as baryon fraction provide key cosmological constraints.

Clusters also provide overwhelming evidence for the existence of
copious amounts of dark matter in the universe. The bulk of the mass
of a cluster is dark matter ($\sim 85 \%$), with hot baryonic gas
contributing about $10\%$ and the rest provided by the stellar content
of the constituent galaxies. 

One of the key predictions of CDM is the excellent fit to density
profiles on a wide range of scales provided by the Navarro-Frenk-White
(NFW) form \citep{NA96.1, NA97.1}, or with modifications to the inner
slope \citep{MO99.2, NA04.1}. In numerical simulations of structure
formation, it is found that this profile fits dwarf galaxy scale dark
matter halos as well as massive, cluster scale dark matter halos. For
a typical cluster dark matter halo in these simulations, the density
profile steepens for radii larger than the halo's typical scale radius
(defined more precisely in \S~\ref{define} below).

A useful diagnostic, the halo concentration, can be defined as the
ratio of the halo's virial radius to its scale radius. This
concentration parameter reflects the central density of the halo,
which depends on the halo's assembly history and thereby on its time
of formation. Since in the hierarchical structure formation scenario
massive galaxy clusters are the most recent bound objects to form,
their concentrations are a crucial probe of the mean density of the
universe at relatively late epochs. 

As originally suggested by \cite{NA96.1} and supported by later
numerical simulations of cosmological structure formation
\citep{BU01.1, HE07.1}, a halo's concentration parameter is related to its
virial mass, with the concentration decreasing gradually with
mass. Given this prediction it is important to test 
this trend with observational data, as this offers an indirect check
on the veracity of the paradigm itself. The situation at present with
observed clusters with measured concentrations is unclear due to the
plurality of methods employed to derive these concentrations as well
as systematics arising from the complex dynamics and its effect on
mass distributions.

In this paper, we present new concentration measurements for 10 strong
lensing clusters and combine our results with other measurements in
the literature to construct an observed concentration-mass relation
for galaxy clusters. In \S~\ref{define}, we present the basic definitions
and equations relevant to computing concentrations. In the following
section, \S~\ref{sample}, we define the observational sample and the
various methods used to select these clusters. We present the
constructed observed concentration-mass relation in \S~\ref{cmrelation}, and
examine the distribution of concentrations at fixed mass in \S~\ref{fixmass}.
In \S~\ref{highc}, we explore the physical effects that might cause
anomalously high measurements of concentration for some clusters, and
present our conclusions in \S~\ref{summary}.

One of the key points we emphasize is that the clusters in this
compiled sample have their masses and concentrations measured using a
variety of different methods: strong lensing, weak lensing, X-ray
temperatures, line of sight velocity distributions, and the caustic
method.  In several cases, a cluster's mass and concentration are
measured using multiple methods that yield different results, and it
is these discrepancies that are of interest. Throughout this paper, we
adopt a spatially flat cosmological model dominated by cold dark
matter and a cosmological constant ($\Omega_\mathrm{m0}=0.3$,
$\Omega_\mathrm{\Lambda0}=0.7$, $h=0.7$). 

\section{Definition of the Concentration Parameter}
\label{define}

Cosmological simulations of structure formation suggest that dark
matter halos, independent of mass or cosmology, follow the density
profile given by \cite{NA96.1, NA97.1}. The spherically averaged
NFW profile is given by

\begin{equation}
  \rho(r)=
  \frac{\rho_\mathrm{s}}{(r/r_\mathrm{s})(1+r/r_\mathrm{s})^2}\;,
\end{equation}
where $\rho_\mathrm{s}$ is a characteristic density and $r_\mathrm{s}$
is the scale radius, which describes the transition point where the
density profile turns over from $\rho\propto r^{-1}$ to $\rho\propto
r^{-3}$. The mass contained within radius $r$ of an NFW halo that
produces gravitational lensing of the background sources is
\begin{equation}
 M(\leq r)=4\pi\Sigma_\mathrm{crit}\kappa_\mathrm{s}r_\mathrm{s}^2\,
  \left[\ln(1+x)-\frac{x}{1+x}\right]\;, 
\end{equation} 
where $x \equiv r/r_\mathrm{s}$ and $\Sigma_\mathrm{crit}$ is
the critical surface mass density, defined as 
\begin{equation}
\Sigma_\mathrm{crit} \equiv \frac{c^2} {4 \pi G} \frac{D_{s}} {D_{l}
  D_{ls}} \; ,
\label{sigmacrit}
\end{equation}
which depends on the angular diameter distances $D_{l,s,ls}$ from the
observer to the lens, to the source, and from the lens to the source,
respectively. The scale convergence is $\kappa_\mathrm{s} =
\rho_\mathrm{s} r_\mathrm{s} / \Sigma_\mathrm{crit}$ \citep{BA96.1}.

The concentration parameter of a halo is the ratio of its virial
radius to its scale radius, and is representative of the halo's
central density. In much of the literature discussing cluster
concentrations, two distinct definitions of the virial radius are
commonly used.  First, the virial radius may be defined as the radius
$r_{200}$ at which the average halo density is 200 times the critical
density at the halo redshift. In this case, the concentration is
denoted as $c_{200} \equiv r_{200}/r_\mathrm{s}$. In an alternative
convention, the virial radius is defined as the radius
$r_\mathrm{vir}$ at which the average halo density is
$\Delta_\mathrm{vir} (z)$ times the mean density at the halo redshift
$z$, where $\Delta_\mathrm{vir} (z)=(18 \pi^2 +82x-39x^2)/(1+x)$ and $x \equiv
\Omega_m(z)-1$ \citep{HU03.1}.  The resultant halo concentration is
$c_\mathrm{vir} \equiv r_\mathrm{vir}/r_\mathrm{s}$.  

For ease of comparison with other work, we will henceforth report all
measurements in terms of both definitions of virial radius: $c_{200}$
and the corresponding halo mass $M_{200} \equiv M(\leq r_{200})$, and
$c_\mathrm{vir}$ and the corresponding $M_\mathrm{vir} \equiv M(\leq
r_\mathrm{vir})$.

\begin{table*}
\begin{center}
\caption[]{New cluster concentrations and masses determined via strong
  lensing.}
\label{tbl-0}
\begin{tabular}{cccccc}
\hline
\hline
Cluster & Lens$^{a}$ & $c_{200}$ & $M_{200}$ &
  $c_\mathrm{vir}$ & $M_\mathrm{vir}$ \\ 
& & & $(10^{14} \, M_\odot)$ & & $(10^{14} \, M_\odot)$ \\ 
\hline

ClG~2244$-$02 & & $4.3 \pm 0.4$ & $4.5 \pm 0.9$ & $5.2
\pm 0.5$ & $5.2 \pm 1.1$  \\
Abell~370 & G1 & $4.8 \pm 0.2$ & $9.0 \pm 1.0$ & $5.8 \pm 0.3$ & $10. \pm 1$ \\
& G2 & $5.2 \pm 0.3$ & $6.7 \pm 0.7$ & $6.3 \pm 0.3$ & $7.7 \pm 0.8$ \\
3C~220.1 & & $4.3 \pm 0.2$ & $3.1 \pm 0.3$ & $5.0 \pm 0.2$ & $3.5 \pm 0.3$ \\
MS~2137.3$-$2353 & & $13 \pm 1$ & $ 2.9 \pm 0.4$ & $16 \pm 1$ &
$3.2 \pm 0.4$ \\ 
MS~0451.6$-$0305 & & $5.5 \pm 0.3$ & $18 \pm 2$ & $6.4 \pm 0.3$ &
$20. \pm 2$ \\ 
MS~1137.5$+$6625 & & $3.3 \pm 0.2$ & $6.5 \pm 0.7$ & $3.8 \pm 0.2$ &
$7.2 \pm 0.8$ \\ 
ClG~0054$-$27$^{b}$ & G1 & $1.2 \pm 0.1$ & $0.42 \pm 0.07$
& $1.5 \pm 0.1$ & $0.52 \pm 0.09$ \\
& G2 & $2.1 \pm 0.1$ & $0.95 \pm 0.12$ & $2.5 \pm 0.1$ & $1.1 \pm 0.1$ \\
Cl~0016$+$1609$^{b}$ & DG 256 & $2.1 \pm 0.1$ & $1.1 \pm
0.2$ & $2.5 \pm 0.1$ & $1.3 \pm 0.2$ \\ 
& DG 251 & $2.3 \pm 0.1$ & $0.51 \pm 0.06$ & $2.7 \pm 0.1$ & $0.59 \pm 0.07$ \\
& DG 224 & $3.1 \pm 0.2$ & $3.1 \pm 0.4$ & $3.6 \pm 0.2$ & $3.6 \pm 0.4$ \\
Cl~0939$+$4713 & G1 & $4.5 \pm 0.3$ & $0.71 \pm 0.11$ & 
$5.4 \pm 0.4$ & $0.81 \pm 0.12$ \\
& G2 & $3.7 \pm 0.3$ & $1.1 \pm 0.2$ & $4.5 \pm 0.3$ & $1.2 \pm 0.2$ \\
& G3 & $4.5 \pm 0.2$ & $1.4 \pm 0.1$ & $5.4 \pm 0.3$ & $1.6 \pm 0.2$ \\
ZwCl~0024$+$1652 & \#362 & $4.6 \pm 0.2$ & $3.1 \pm 0.1$ & $5.5 \pm
0.3$ & $2.6 \pm 0.3$ \\  
& \#374 & $4.3 \pm 0.2$ & $3.7 \pm 0.5$ & $5.1 \pm 0.3$ & $4.2 \pm 0.5$ \\
& \#380 & $3.4 \pm 0.2$ & $2.7 \pm 0.3$ & $4.1 \pm 0.2$ & $3.2 \pm 0.3$ \\

\hline
\end{tabular}
\flushleft$^a$See \cite{CO06.1} for lens identification.
\vspace{-2mm} \flushleft $^b$Because these cluster arcs have no published
  redshifts, we calculate the concentration and mass assuming
  $D_s/D_{ls}=1$. Note that we do not use these concentration and mass
  determinations in our observed concentration-mass fit.
\end{center}
\end{table*}

\section{Observational Sample}
\label{sample}

To determine the observed relation between concentration and virial
mass for galaxy clusters, we compile a sample of all known
observationally-determined concentrations and the corresponding virial
masses.  The bulk of our sample is drawn from pre-existing data in the
literature, but we also incorporate new concentration and mass
determinations for 10 strong lensing clusters. This compilation
presents the most complete sample of observed cluster concentrations
and virial masses yet assembled; in total, our sample consists of 182
unique measurements for 100 galaxy clusters.

\subsection{New Concentrations for 10 Strong Lensing Clusters}
\label{new}

\cite{CO06.1} used the strong lensing arcs observed in 10 galaxy
clusters to fit elliptical NFW dark matter density profiles to each
cluster.  Using their best-fit scale radius and scale convergence
parameters, as well as the observed cluster and arc redshifts, we
determine a concentration and mass for each lens, shown in
Table~\ref{tbl-0}.  We compute errors in concentration and mass based
on the errors derived for the best-fit NFW parameters. As detailed in
\cite{CO06.1} these errors are quite small (and likely underestimate
the true error) because they apply to one particular model and do not
reflect degeneracies between models.

In particular, for several lensing clusters, the best-fit mass
distribution is bi-modal and due to the difficulty of converting
these mass models accurately to a single NFW parameterization to
derive the concentration, we retain in our sample only clusters
that are well defined by a primary single dark matter halo. Our fits
to the clusters ClG~2244$-$02, 3C~220.1, MS~2137.3$-$2353,
MS~0451.6$-$0305, and MS~1137.5$+$6625 fulfill this criterion, more
than doubling the number of existing cluster concentration and mass
measurements, with errorbars, from strong lensing. 

Two of the clusters from \cite{CO06.1}, ClG~0054$-$27 and
Cl~0016$+$1609, do not have published arc redshifts, preventing us
from determining the cluster's critical surface mass density
(Equation~\ref{sigmacrit}) and therefore the virial mass or the
concentration parameter. Instead, we assume a ratio of angular
diameter distances $D_s/D_{ls}=1$ to calculate the concentrations and
masses for these two clusters, reported in Table~\ref{tbl-0}.  We
exclude these two clusters, however, from the analysis that follows.

\subsection{Compiling Published Observations}
\label{old}

We combine our concentration and mass measurements with those in the
literature to create a complete sample of observed galaxy cluster
concentrations and virial masses. The mass distributions of these
clusters are fit using a variety of observational methods, and in many
cases a cluster is independently fit by different authors using
different methods. The assumption of spherical or axial symmetry of
the halo often figures prominently in mass calculations.  We briefly
outline below each of the methods employed in the determination of a
dark matter halo's mass distribution.  

{\it Strong lensing (SL)} --- Strong lensing typically occurs when the
projected surface mass density in the inner regions of a cluster 
is sufficiently high to produce one or more distorted images of a
single background galaxy. The observed positions, orientations, and
magnifications of the lensed images are used to constrain the mass 
distribution of the cluster lens. The integrated mass determined in
the inner regions of a cluster from strong lensing effects is 
often systematically higher than that determined using X-ray data, 
leading to higher values of the estimated concentration parameter 
compared to the average galaxy cluster that is not a lens. This effect
is due to the fact that lensing clusters and in particular those that
exhibit strong lensing tend to preferentially sample the high mass end
of the cluster mass function.

{\it Weak lensing (WL)} --- The gravitational tidal field caused by a mass
distribution produces elongated, tangential distortions of background
objects. These weak distortions are observed at large radii from
the center of lensing clusters and their statistical analyses provide
a direct measure of the density profile of the cluster lens at
intermediate to large radii.  

{\it Combined weak lensing and strong lensing (WL+SL)} --- If both
weak and strong lensing measurements of a cluster are used in
combination, they can be used to effectively break
the mass-sheet degeneracy \citep{SC95.1} that often plagues solitary
strong or weak lensing mass measurements. Therefore, the combination
of strong and weak lensing can provide a more accurate and calibrated
mass distribution for a cluster.

{\it X-ray temperature} --- The hot gas in galaxy clusters emits X-rays via
bremsstrahlung radiation and atomic line emission. The surface
brightness distribution and the measured cluster temperature
can be used to determine the density profile. Combining the
temperature and density information yields the cluster mass.  
However, the X-ray technique for mass determination assumes that 
the intra-cluster gas is distributed in a spherically symmetric 
fashion and is in hydrostatic equilibrium \citep{EV96.1}. These
assumptions may be untenable; for example, observed buoyant 
bubbles near the cores of galaxy clusters suggest the hot gas 
may not be strictly in hydrostatic equilibrium (e.g., \citealt{CH01.1}). 

{\it Line of sight velocity distribution (LOSVD)} --- The LOSVD for a
cluster is a function relating the number of cluster galaxies with their
line of sight velocities. This function is parameterized by 
velocity moments, which can be measured observationally (for example,
the second moment of the LOSVD is the square of the line of sight
velocity dispersion). Combining these results with the Jeans equation,
assuming spherical symmetry, determines the potential and therefore the 
density profile of the cluster itself.

{\it Caustic method (CM)} --- When the line of sight velocity is
plotted against the projected clustercentric radius, member galaxies
in a cluster align in a distinctively flaring pattern. The edges of
these flares are called caustics and demarcate the cluster infall
region \citep{KA87.1}. Based on the location and amplitude of these
caustics, we can infer the cluster potential and therefore the mass of
a cluster. However, this technique relies upon the assumptions of
spherical symmetry and does not adequately take into account the
non-linearities in structure formation.

A variety of definitions are used in the literature for the
virial radius, and for consistency we employ the \cite{HU03.1} formula
to convert all concentrations and masses to our preferred convention 
of ($c_{200}$, $M_{200}$) and ($c_\mathrm{vir}$, $M_\mathrm{vir}$). 
We also convert all the data to a flat $\Lambda$CDM cosmological 
model ($\Omega_\mathrm{m0}=0.3$, $\Omega_\mathrm{\Lambda0}=0.7$,
$h=0.7$). The entire sample is presented in Table~\ref{tbl-1}, which
we use to define an observed cluster concentration-mass relation.

\section{The Observed Concentration-Mass Relation}
\label{cmrelation}

To discern the trend between concentration and mass, we first cull our
sample down to only those concentrations that have corresponding
virial mass estimates and that have published errors in both
quantities. This narrows our sample down to 62 clusters. 

Several clusters have multiple, distinct measurements of their
concentration and virial mass, leading them to be over-represented in
our sample in comparison to clusters with a single measurement. To
address this problem, we take the median value of the concentration
and its corresponding virial mass to be representative of a cluster
with multiple measurements. This is done consistently in our analysis.

\begin{figure}
\includegraphics[angle=0,width=80mm]{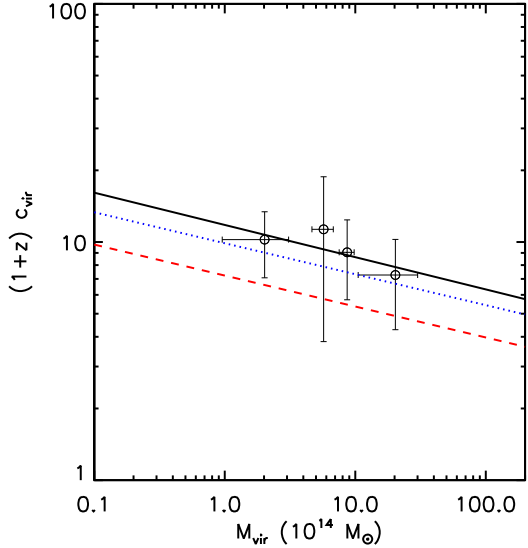}
\caption{Observed cluster concentrations and virial masses, binned by
  mass with approximately equal numbers of clusters in each bin.
  Data points illustrate the mean concentration and mass of each bin, and the
  solid line is our best fit. The slope of our fit is consistent
  with fits from simulations, \protect\cite{HE07.1} (dotted blue line) and
  \protect\cite{BU01.1} (dashed red line), though our normalization is
  somewhat higher.}  
\label{fig:cbin}
\end{figure}

\begin{figure}
\includegraphics[angle=0,width=80mm]{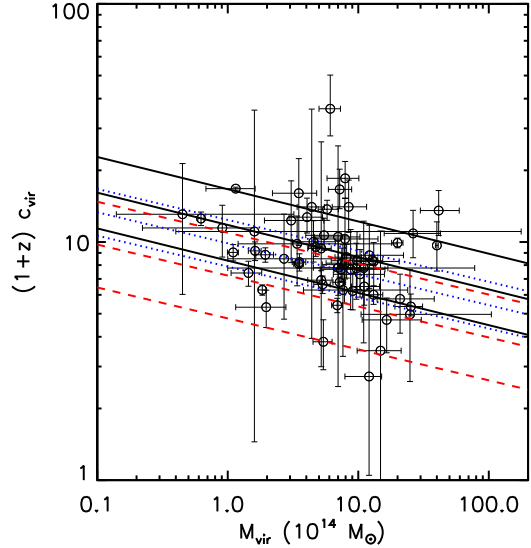}
\caption{Observed cluster concentrations and virial masses, as well as
the best-fit power law $c_\mathrm{vir}=\frac{14.8} {(1+z)}
(M_\mathrm{vir}/M_\star)^{-0.14}$ (solid black line).  The outer two
solid black lines depict the 1-$\sigma$ scatter $\Delta(\log
c_\mathrm{vir}) \sim 0.15$.  Also plotted are two $c-M$ relations from
simulated clusters: the \protect\cite{HE07.1} $c_\mathrm{vir}=\frac{12.3}
{(1+z)} (M_\mathrm{vir}/M_\star)^{-0.13}$ (dotted blue line, with outer 
dotted blue lines as 1-$\sigma$ scatter) and the \protect\cite{BU01.1}
$c_\mathrm{vir}=\frac{9} {(1+z)} (M_\mathrm{vir}/M_\star)^{-0.13}$
(dashed red line, with outer dashed red lines as 1-$\sigma$ scatter). }   
\label{fig:cmvirz0med}
\end{figure}

\begin{figure}
\includegraphics[angle=0,width=80mm]{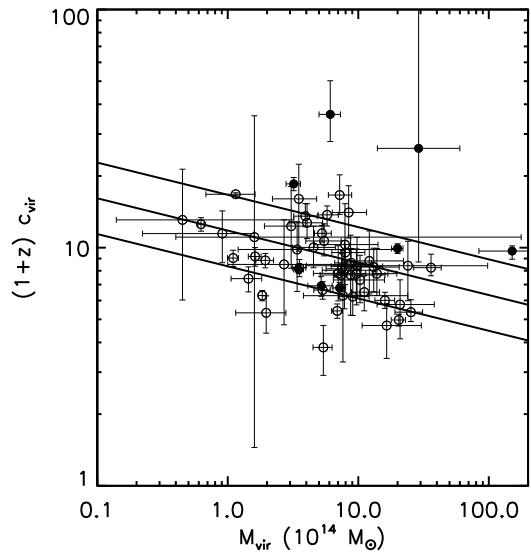}
\caption{Observed cluster concentrations and virial masses derived
  from lensing (filled circles) and X-ray (open circles)
  measurements.  For reference, the solids lines depict the best-fit
  power law to our complete sample and its 1-$\sigma$ scatter. The
  lensing concentrations appear systematically higher than the X-ray
  concentrations, and a Kolmogorov-Smirnov test confirms that the
  lensing results likely belong to a different distribution. }
\label{fig:cmvirlensx}
\end{figure}

As the distribution of cluster concentrations is so broad, we bin the
data into four mass bins with approximately equal numbers of clusters in
each bin for a more effective comparison. We
determine the mean and standard deviation of the cluster
concentrations (normalized to $z=0$) and masses in each bin, then fit
a power law,

\begin{equation} 
c_\mathrm{vir} = \frac{c_0} {1+z} \left( \frac{M_\mathrm{vir}} {M_\star} \right)^\alpha \; ,
\end{equation}
where $c_0$ and $\alpha$ are constants, $z$ is the cluster redshift,
and the mass normalization is taken to be $M_\star=1.3 \times 10^{13}
h^{-1} M_\odot$ as in the simulations. The best fit we obtain is
$c_\mathrm{vir}=\frac{14.8 \pm 6.1} {(1+z)}
(M_\mathrm{vir}/M_\star)^{-0.14 \pm 0.12}$,
shown in Figure~\ref{fig:cbin}.  We compare our fit to the $c-M$
relations inferred from dissipationless $N$-body simulations of
$\Lambda$CDM cosmic structure formation, which are $c_\mathrm{vir}=\frac{9}
{(1+z)} (M_\mathrm{vir}/M_\star)^{-0.13}$ from \cite{BU01.1} and
$c_\mathrm{vir}=\frac{12.3} {(1+z)} (M_\mathrm{vir}/M_\star)^{-0.13}$
from \cite{HE07.1}.  

Our fit has approximately the same slope as the simulations, but a
somewhat larger normalization by a factor of 1.6 compared to
\cite{BU01.1} and a factor of 1.2 compared to \cite{HE07.1}
simulations. 

Our finding of a slope $\alpha$ consistent with simulations is significant,
because previous studies of observed clusters have not found this
agreement.  Fits to X-ray clusters of mass $> 10^{14} M_\odot$ have
found $\alpha \sim 0$, or an approximately constant concentration-mass
relation \citep{PO05.1, VI06.1}.  In the opposite extreme, studies of
X-ray galaxy groups and poor clusters with lower virial masses in the 
range of $\sim 10^{13}\,M_{\odot}$ find steep slopes of 
$\alpha=-0.226$ \citep{GA06.1} and $\alpha=-0.44$ \citep{SA00.1}.
Finally, the \cite{BU06.1} sample of X-ray galaxy systems ranging in
mass from $10^{13} M_\odot$ to $10^{15} M_\odot$ has a slope of
$\alpha=-0.172$. 

Both observations and simulations find a large scatter in
concentration for a given virial mass, which is likely due to the
variation in halo collapse epochs and histories \citep{BU01.1}.
Comparing our best fit to the unbinned clusters, shown in
Figure~\ref{fig:cmvirz0med}, we find a 1-$\sigma$ scatter of
$\Delta(\log c_\mathrm{vir}) \sim 0.15$ in our relation.  The
simulations have scatters of $\sim 0.18$ in \cite{BU01.1} and $\sim
0.098$ in \cite{HE07.1} (calculated from data courtesy of
J. Hennawi).  However, we cannot directly compare these to the
observationally derived scatters due to the differing systematics. 

\begin{figure}
\includegraphics[angle=0,width=80mm]{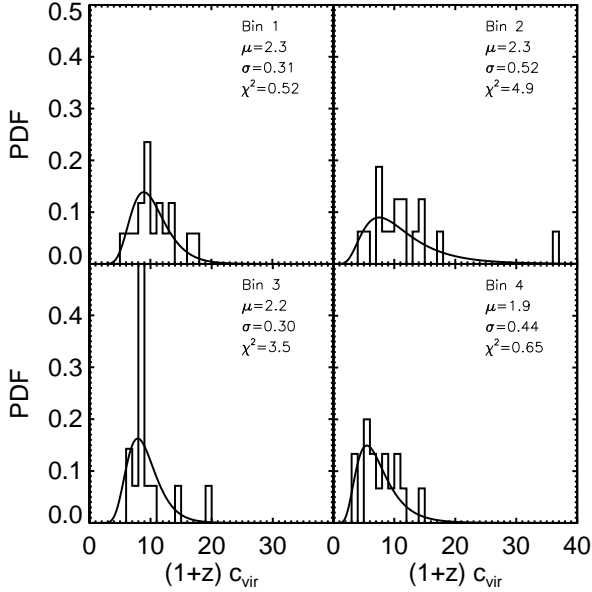}
\caption{Log-normal fits to normalized histograms of observed cluster
  concentrations, binned by mass.  The mass ranges are $M_\mathrm{vir}
  < 4 \times 10^{14} M_\odot$, $4 \times 10^{14} M_\odot <
  M_\mathrm{vir} < 7.3 \times 10^{14} M_\odot$, $7.3 \times 10^{14}
  M_\odot < M_\mathrm{vir} < 12 \times 10^{14} M_\odot$, and
  $M_\mathrm{vir} > 12 \times 10^{14} M_\odot$ for bin 1 to bin 4,
  respectively. For each mass bin the expectation
  value $\mu$, standard deviation $\sigma$, and $\chi^2$ of the
  best-fit log-normal function are also given.  } 
\label{fig:lognorm}
\end{figure}

We note that the concentrations of clusters determined from lensing
methods (weak, strong, and a combination of the two) are
systematically higher than the concentrations determined by other
methods. Figure~\ref{fig:cmvirlensx} shows the distribution of
lensing concentrations relative to X-ray concentrations; a
Kolmogorov-Smirnov test finds only a 28\% probability that the two are
in fact derived from the same parent distribution.

A similar, though less pronounced, effect has been found in numerical
simulations of clusters. \cite{HE07.1} identified lensing clusters
from their simulated sample by using ray tracing to compute strong
lensing cross sections for each cluster, and found that the simulated
strong lensing clusters have on average 34\% higher concentrations
than the total simulated cluster population.  In our observed sample,
we find a larger fraction: about 55\% higher concentrations on average
for observed strong lensing clusters.

Why are lensing concentrations systematically higher than X-ray
concentrations? Several known physical effects are implicated in
explaining this discrepancy.  The X-ray method of determining a mass
distribution depends on the assumption of hydrostatic equilibrium,
which breaks down for unrelaxed clusters. As a result, X-ray
measurements may underpredict concentrations for unrelaxed systems
such as clusters undergoing mergers. 

In particular, if non-thermal sources of pressure support are present
and significant, for example due to the presence of a magnetic field
on small scales in the inner regions of a cluster, the assumption of
hydrostatic equilibrium will tend to underestimate the total mass
and hence yield a systematically lower value for the
concentration. \cite{LO94.1} argue that for the cluster Abell~2218, a
factor of about $2 - 3$ discrepancy in the strong lensing determined mass
and the X-ray determined mass (under the assumption of hydrostatic
equilibrium) enclosed within 200 kpc can be explained with the existence
of an equi-partition magnetic field.  This effect alone could likely
close the gap between X-ray and lensing concentrations.

In addition to underpredicted X-ray concentrations, overpredicted lensing
concentrations could also contribute to the discrepancy in
concentration estimates.  In particular, lensing concentrations can be
inflated due to the effects of halo triaxiality, substructure along
the line of sight, and adiabatic contraction in the dark matter due to
the collapse of baryons in the inner regions of halos.

\begin{figure}
\includegraphics[angle=0,width=80mm]{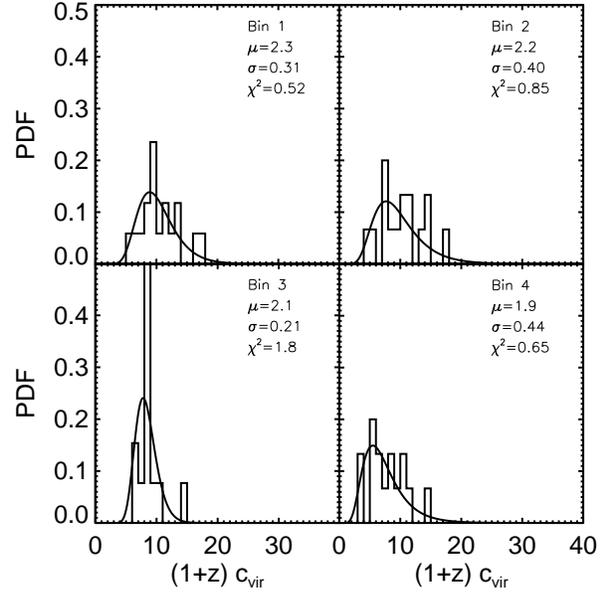}
\caption{As Fig.~\ref{fig:lognorm}, but with concentrations greater
  than 2-$\sigma$ from the expectation value omitted.  This excluded
  the highest concentration cluster in bin 2 and in bin 3.  The
  resultant fits improved by up to a factor of 6 in $\chi^2$.} 
\label{fig:lognorm_no_outlier}
\end{figure}

It is impossible to observationally determine a distant halo's
three-dimensional shape, and most mass-finding techniques assume a
spherical halo.  However, a spherical halo model fit to a triaxial
cluster, if projected along the major axis, would overestimate both
the cluster's concentration and its virial mass \citep{GA05.2,
OG05.1}.  If a halo were significantly elongated along the line of
sight, its concentration could be overestimated up to 50\% and its
virial mass estimation could double \citep{CO06.2}.  Methods now exist
to estimate the shape of a dark matter halo from the observed
intracluster gas \citep{LE03.2}.

Structure along the line of sight to the cluster can also contribute
to a higher estimated concentration. Simulations of \cite{KI07.1}
determined that multiple subhalos close to the line of sight are most
effective at increasing the concentration estimate of the main halo.
Neglecting large-scale structure, as most halo mass models do, can  
artificially and substantially inflate concentration estimates.

Finally, adiabatic contraction in the halo core could substantially
increase a cluster's concentration, as argued by \cite{GN04.1}.  The
dissipative collapse of baryons in the centers of dark matter halos
induces a steepening of the dark matter density profile in these
regions. This steepening will systematically increase the
concentrations.  Adiabatic contraction could explain why our observed
lensing concentrations are yet higher than the simulated cluster lens
concentrations in our comparison.  The \cite{HE07.1} simulated
lensing clusters are products of dissipationless simulations, whereas
observed clusters have presumably undergone adiabatic contraction and
a corresponding steepening in the density profile, yielding a higher
value for the concentration.

\begin{figure}
\includegraphics[angle=0,width=80mm]{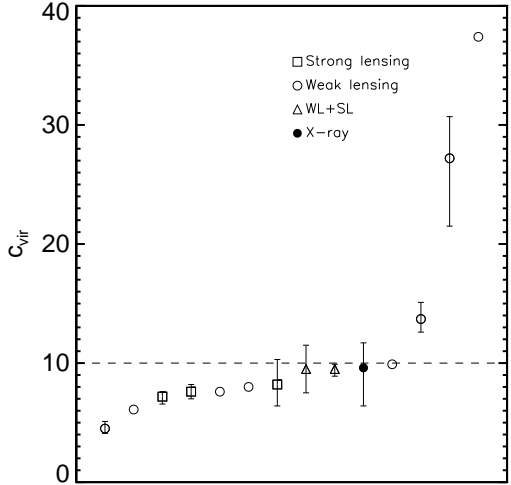}
\caption{Observed concentrations for the cluster Abell 1689, ordered
  from lowest to highest.  Only weak lensing measurements produce the
  anomalously high $c_\mathrm{vir} > 10$ concentrations (dashed line). }
\label{fig:a1689}
\end{figure}

\section{Concentrations for Fixed Halo Mass}
\label{fixmass}

Numerical simulations further indicate that concentrations for fixed
halo mass are log-normally distributed. To test this hypothesis for
observed clusters, we examine the clusters grouped into four mass
bins, as detailed in \S~\ref{cmrelation}.  We then fit a log-normal
function to the distribution of concentrations in each bin.  For our
$x=(1+z) c_\mathrm{vir}$, the log-normal probability density function
(PDF) is 

\begin{equation}
f(x;\mu,\sigma)=\frac{1} {\sqrt{2 \pi} \sigma x} \exp \left(
  \frac{-(\ln{x}-\mu)^2} {2 \sigma^2} \right) \; ,
\end{equation}
where $\mu$ and $\sigma$ are the expectation value and standard
deviation.

The panels in Figure~\ref{fig:lognorm} show the best-fit log-normal
functions to each mass bin, as well as the expectation value, standard
deviation, and $\chi^2$ for each fit.  The concentrations appear to be
consistent with a log-normal distribution, with the exception of a
couple of high concentration clusters that lie beyond the tail of the
distribution.  To determine how well the bulk of concentrations,
without the outliers, is fit by a log-normal distribution, we omit
all concentrations greater than 2-$\sigma$ from the expectation value.
This cut eliminates one cluster from bin 2 (top row, right in
Fig.~\ref{fig:lognorm}) and one from bin 3 (bottom row, left in
Fig.~\ref{fig:lognorm}). 

The resultant fits, shown in Figure~\ref{fig:lognorm_no_outlier}, have
improved in goodness of fit to the data by a factors of 6 and 2 in
$\chi^2$ for bin 2 and bin 3, respectively.  All four bins are now
well-fit by a log-normal function, suggesting that the vast majority
of observed clusters follow a log-normal distribution, but with a few
outliers with substantially higher concentrations.

These outliers are ZwCl~0024$+$1652 (in bin 2) and MS~2137.3$-$2353
(in bin 3), and their anomalously high concentrations are well
documented in the literature.  Possible explanations for these high
concentrations are presented in the next section.

\begin{figure}
\includegraphics[angle=0,width=80mm]{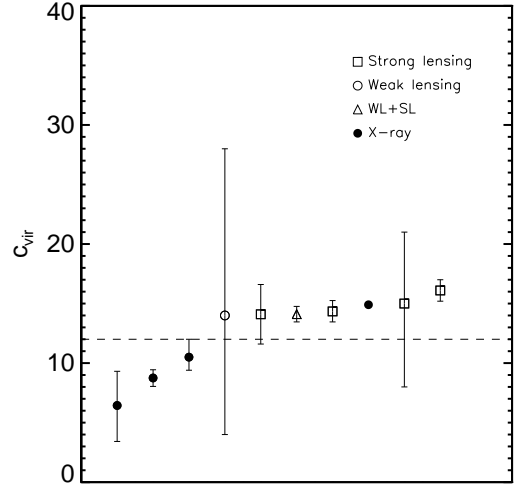}
\caption{Observed concentrations for the cluster
  MS~2137.3$-$2353, ordered from lowest to highest. Note that all of
  the smaller values of concentration $c_\mathrm{vir} < 12$ (dashed
  line), which are more consistent with predictions from observed and
  simulated $c-M$ relations, are from X-ray measurements.}  
\label{fig:ms2137}
\end{figure}

\section{Anomalously High Concentration Clusters}
\label{highc}

Three clusters stand out for their anomalously high concentrations,
which are several sigma higher than the predicted concentration-mass
relations.  They are the lensing clusters ZwCl~0024$+$1652
($c_\mathrm{vir}=26$; \citealt{KN03.1}), Abell~1689, and
MS~2137.3$-$2353.

Although there are only lensing measurements of the ZwCl~0024$+$1652
concentration, multiple strong lensing, weak lensing, combined strong
and weak lensing, and X-ray measurements have been made of the
concentrations and virial masses of both Abell~1689 and
MS~2137.3$-$2353.  Many of these individual measurements are even
consistent with concentration-mass relations, leading us to ask
whether the anomalously high concentrations are indeed real, and what
the physical explanation for these high concentrations might be.

Figure~\ref{fig:a1689} illustrates the range of concentrations
measured for the cluster Abell~1689. Strikingly, if not for the
weak lensing measurements, this cluster would have a rather typical
range of concentrations.  Apparently some systematic effect causes the
weak lensing concentrations to be inflated relative to concentrations
inferred via other methods.  For instance, the weak lensing signal
could be more sensitive to substructure close to the line of sight,
causing concentration overestimates, as discussed in
\S~\ref{cmrelation}.

We see an analogous effect in MS~2137.3$-$2353, shown in
Figure~\ref{fig:ms2137}.  Here, all of the smaller, unsurprising
concentration values are the result of X-ray measurements.  Lensing
measurements produce most of the anomalously high concentrations.

As discussed in \S~\ref{cmrelation}, X-ray concentrations can be
systematically low for clusters that do not conform to the assumption
of hydrostatic equilibrium. And, lensing concentrations can be elevated
due to projection effects, substructure, or adiabatic contraction.  We
expect lensing concentrations to be greater than X-ray concentrations,
but it is yet unclear whether the anomalously high lensing
concentrations of these three clusters are real.

This concern has been addressed for each of the three anomalously high
concentration clusters. The high lensing concentration of
MS~2137.3$-$2353 may be explained by an elongated halo with its major
axis close to the line of sight \citep{GA05.2}. Also, ZwCl~0024$+$1652
exhibits prominent substructure that may account for its high
concentration estimates.  For example, \cite{KN03.1} identify a main
clump with high concentration as well as a secondary, low
concentration clump, while this paper (Table~\ref{tbl-0}) fits three
$c_\mathrm{vir} \sim 5$ clumps to ZwCl~0024$+$1652.

Perhaps the most progress has been made in explaining Abell~1689's
concentration, which has been measured to be as high as $c_\mathrm{vir}=37.4$
\citep{HA06.1} but has come down to a consensus of $c_\mathrm{vir}
\sim 6 - 8$ \citep{LI06.1} due to careful, detailed modeling including
an unprecedented large number of strong lensing constraints.  

\section{Summary and Conclusions}
\label{summary}

We have presented a comprehensive set of observed galaxy cluster
concentrations and virial masses, including new concentration
estimates for 10 strong lensing clusters.  With this data, we fit the
dependence of the concentration parameter with virial mass to a power
law to compare with the relation obtained in simulations. The main results
of this analysis are:

1. The observed cluster concentrations and virial masses are best fit
   by the power law $c_\mathrm{vir}=\frac{14.8 \pm 6.1} {(1+z)}
   (M_\mathrm{vir}/M_\star)^{-0.14 \pm 0.12}$ with $M_\star = 1.3 \times
   10^{13}\,h^{-1}\,M_{\odot}$.  The slope is consistent with the
   value of $-0.13$ found by simulations, in contrast to previous
   observational studies which found a steeper slope or no slope at
   all.  The normalization of our best fit is at least 20\% higher
   than the normalizations found by simulations. We suspect that adiabatic
   contraction and a steepening of the dark matter density profile in
   response to the collapse of baryons in real clusters offers a likely 
   explanation for this systematic offset.       

2. Cluster concentrations derived from lensing analyses are
   systematically higher than concentrations derived via X-ray
   temperatures. We find that observed strong lensing clusters have
   concentrations 55\% higher, on average, than the rest of the
   cluster population, a larger factor than found in simulations.  The
   discrepancy between lensing and X-ray concentrations is likely due
   to some combination of X-ray concentrations underpredicted for
   unrelaxed clusters and lensing concentrations overpredicted due to
   halo triaxiality, structure along the line of sight, and adiabatic
   contraction. 

3. For fixed mass, the majority of observed clusters are distributed
   log-normally in concentration, with a few exceptions.  The
   log-normal distribution is predicted by simulations, but has not
   been measured observationally prior to this work. The exceptions to
   this log-normal distribution are two clusters well-known for their
   anomalously high concentration measurements.

4. The three clusters with the highest concentration measurements have
   been well studied, and the physical effects (such as halo
   elongation, substructure, and adiabatic contraction) behind these large
   concentrations are better understood.  These effects need to 
   be accounted for with careful modeling.

Although our observed concentration-mass relation for galaxy clusters
is reasonably consistent (albeit with a higher normalization) with
present simulations, the Millennium simulation will offer the best
comparison set to the observed relation reported here. This simulation
will offer ample statistics spanning all four mass bins for a direct
comparison with our data.

\section*{Acknowledgments}{J.M.C. and P.N. acknowledge insightful and
  useful discussions with Matthias Bartelmann. We also benefited from
  exchanges at the concentration discussion group at the KITP workshop
  on applications of gravitational lensing. J.M.C. acknowledges
  support by a National Science Foundation Graduate Research Fellowship.} 

\bibliographystyle{mn2e}
\bibliography{cmrelation}

\begin{appendix}

\renewcommand{\thetable}{A-\arabic{table}}
\setcounter{table}{0}

\section*{Appendix A: Complete Compilation of Observed Cluster
  Concentrations and Virial Masses}
Table~\ref{tbl-1} contains the full data set of observed cluster
concentrations and virial masses used in this paper.  We convert all
concentrations and masses to our definitions of virial radius (given
in \S~\ref{define}) and use $h=0.7$ throughout.  In all, there are 182
unique measurements of 100 clusters.

\onecolumn

\begin{center}
\begin{longtable}{llllllll}
\caption{Cluster concentrations and masses} \\
\hline \hline \\[-2ex]
   \multicolumn{1}{c}{Cluster} &
   \multicolumn{1}{c}{$z$} &
   \multicolumn{1}{c}{Method} &
   \multicolumn{1}{c}{$c_{200}$} &
   \multicolumn{1}{c}{$M_{200}$} &
   \multicolumn{1}{c}{$c_\mathrm{vir}$} &
   \multicolumn{1}{c}{$M_\mathrm{vir}$} &
   \multicolumn{1}{c}{Reference} \\ & & & &
\multicolumn{1}{c}{$(10^{14} \, M_\odot)$} & &
\multicolumn{1}{c}{$(10^{14} \, M_\odot)$} & \\ [0.5ex] \hline
   \\[-1.8ex]
\endfirsthead

\multicolumn{8}{c}{{\tablename} \thetable{} -- {\it Continued}} \\[0.5ex]
  \hline \hline \\[-2ex]
   \multicolumn{1}{c}{Cluster} &
   \multicolumn{1}{c}{$z$} &
   \multicolumn{1}{c}{Method} &
   \multicolumn{1}{c}{$c_{200}$} &
   \multicolumn{1}{c}{$M_{200}$} &
   \multicolumn{1}{c}{$c_\mathrm{vir}$} &
   \multicolumn{1}{c}{$M_\mathrm{vir}$} &
   \multicolumn{1}{c}{Reference} \\ & & & &
\multicolumn{1}{c}{$(10^{14} \, M_\odot)$} & &
\multicolumn{1}{c}{$(10^{14} \, M_\odot)$} & \\[0.5ex] \hline
  \\[-1.8ex]
\endhead


\\[-1.8ex] \hline 
\endlastfoot

Virgo & 0.003 & X-ray & $2.8 \pm 0.7$ & $4.2 \pm 0.5$ & $3.8 \pm 0.9$ &
$5.4 \pm 0.9$ & \cite{MC99.1} \\
Abell 1060 & 0.01 & LOSVD & $10.6^{+17.1}_{-7.7}$ & $3.8^{+0.4}_{-0.7}$
& $13.9^{+21.9}_{-10.0}$ & $4.4^{+1.1}_{-1.0}$ & \cite{LO06.1} \\  
 & & X-ray & $8.4 \pm 0.6$ & & $11.1 \pm 0.8$ & & \cite{XU01.1} \\
Abell 262 & 0.0163 & LOSVD & $3.1^{+8.7}_{-2.4}$ &
$2.1^{+0.2}_{-0.6}$ & $4.2^{+11.2}_{-3.2}$ & $2.7^{+1.2}_{-1.0}$ & 
\cite{LO06.1} \\ 
 & & X-ray & $6.7 \pm 0.5$ & $0.929 \pm 0.082$ & $8.9
\pm 0.7$ & $1.100 \pm 0.106$ & \cite{GA06.1} \\
 & & X-ray & $5.29 \pm 0.43$ & & $7.03 \pm 0.55$ & & \cite{VI05.1} \\
 & & X-ray & $12.9 \pm 1.1$ & & $16.8 \pm 1.4$ & & \cite{XU01.1} \\
Abell 194 & 0.018 & CM & $6.27$ & $1.09$ & $8.30$ & $1.30$
& \cite{RI03.1} \\ 
MKW 4 & 0.0200 & X-ray & $9.4 \pm 0.7$ & $0.54 \pm 0.027$ & $12.3 \pm
0.8$ & $0.624 \pm 0.034$ & \cite{GA06.1} \\
 & & X-ray & $3.85 \pm 0.22$ & $1.11 \pm 0.15$ & $5.17 \pm 0.28$
 & $1.37 \pm 0.20$ & \cite{VI05.1} \\
Abell 3581 & 0.0218 & X-ray & $9.81^{+6.30}_{-5.40}$ &
$0.39^{+2.23}_{-0.27}$ & $12.8^{+8.1}_{-6.9}$ &
$0.45^{+2.76}_{-0.31}$ & \cite{VO06.1} \\ 
Abell 1367 & 0.022 & CM & $16.9$ & $5.46$ & $21.9$ & $6.11$ &
\cite{RI03.1} \\ 
Abell 1656 & 0.023 & CM & $10.0$ & $11.2$ & $13.1$ & $12.9$ &
\cite{RI03.1} \\
 & & LOSVD & $7.0$ & $11.8 \pm 0.3$ & $9.3$ & $13.9 \pm 4$ &
 \cite{LO03.1} \\ 
Abell 539 & 0.029 & CM & $14.7$ & $3.63$ & $19.0$ & $4.09$ &
\cite{RI03.1} \\ 
Abell 2199 & 0.030 & CM & $7.47$ & $4.67$ & $9.80$ & $5.47$ &
\cite{RI03.1} \\ 
 & & LOSVD & $7.79^{+11.26}_{-6.02}$ & $6.0^{+1.5}_{-1.8}$ &
 $10.2^{+14.4}_{-7.8}$ & $7.0^{+3.3}_{-2.4}$ & \cite{LO06.1} \\
 & & LOSVD & $4$ & $5$ & $5$ & $6$ & \cite{KE02.2} \\  
 & & X-ray & $8.2 \pm 0.4$ & & $10.7 \pm 0.5$ & & \cite{XU01.1} \\
 & & X-ray & $10$ & & $13$ & & \cite{MA99.1} \\
AWM 4 & 0.0317 & X-ray & $6.8 \pm 0.6$ & $1.375 \pm 0.146$ & $8.9 \pm
0.8$ & $1.619 \pm 0.182$ & \cite{GA06.1} \\
Abell 496 & 0.0329 & CM & $14.0$ & $3.13$ & $18.1$ & $3.53$ &
\cite{RI03.1} \\ 
 & & LOSVD & $6.9^{+12.9}_{-4.8}$ & $4.5^{+0.3}_{-0.7}$ &
 $9.1^{+16.4}_{-6.2}$ & $5.2^{+1.0}_{-1.1}$ & \cite{LO06.1} \\   
 & & X-ray & $10.4 \pm 0.6$ & & $13.5 \pm 0.8$ & & \cite{XU01.1} \\
 & & X-ray & $6$ & & $8$ & & \cite{MA99.1} \\
Abell 2063 & 0.0337 & X-ray & $5.1 \pm 0.3$ & & $6.8 \pm 0.4$ & &
\cite{XU01.1} \\ 
2A 0335+096 & 0.0347 & X-ray & $8.18^{+18.83}_{-7.20}$ &
$1.4^{+115.5}_{-1.0}$ & $10.7^{+23.9}_{-9.3}$  &
$1.6^{+175.4}_{-1.2}$ & \cite{VO06.1} \\  
Abell 2052 & 0.0348 & X-ray & $9.7 \pm 0.7$ & & $12.6 \pm 0.9$ & &
\cite{XU01.1} \\ 
MKW 9 & 0.0382 & X-ray & $5.41 \pm 0.67$ & $1.20 \pm 0.30$ & $7.14 \pm 0.86$ &
$1.44 \pm 0.38$ & \cite{PO05.1} \\  
 & & X-ray & $5.4 \pm 0.7$ & $1.20$ & $7.1 \pm 0.9$
& $1.44$ &  \cite{PR05.1} \\  
Abell 3571 & 0.039 & X-ray & $4.9 \pm 0.2$ & & $6.5 \pm 0.3$ & &
\cite{XU01.1} \\ 
Abell 576 & 0.04 & CM & $10.9$ & $9.51$ & $14.1$ & $10.85$ &
 \cite{RI03.1} \\ 
RXJ0137 & 0.0409 & X-ray & $6.34$ & $0.99$ & $8.32$ & $1.17$ &
\cite{RI06.1} \\ 
 & & X-ray & $4.9 \pm 2.4$ & & $6.5 \pm 3.1$ & & \cite{BU04.1} \\
Abell 160 & 0.0432 & X-ray & $10.14$ & $0.91$ & $13.16$ & $1.04$ &
\cite{RI06.1} \\ 
Abell 1983 & 0.0442 & X-ray & $3.83 \pm 0.71$ & $1.59 \pm 0.61$ &
$5.10 \pm 0.91$ & $1.97 \pm 0.82$ & \cite{PO05.1} \\    
Abell 119 & 0.0446 & CM & $6.29$ & $4.07$ & $8.25$ & $4.81$ &
\cite{RI03.1} \\ 
 & & X-ray & $2.55$ & $2.36$ & $3.45$ & $3.06$ & \cite{RI06.1} \\
 & & X-ray & $3.3 \pm 0.2$ & & $4.4 \pm 0.3$ & &
\cite{XU01.1} \\ 
MKW 3S & 0.045 & X-ray & $6.4 \pm 0.7$ & & $8.4 \pm 0.9$ & &
\cite{XU01.1} \\ 
Abell 168 & 0.0451 & CM & $5.19$ & $4.30$ & $6.84$ & $5.17$ &
\cite{RI03.1} \\ 
 & & X-ray & $7.69$ & $2.24$ & $10.03$ & $2.61$ & \cite{RI06.1} \\
Abell 4059 & 0.0478 & X-ray & $4.8 \pm 0.2$ & & $6.3 \pm 0.3$ & &
\cite{XU01.1} \\ 
Abell 3558 & 0.048 & LOSVD & $1.9^{+4.0}_{-1.2}$ & $9.0^{+0.3}_{-2.3}$ &
$2.6^{+5.1}_{-1.6}$ & $12.1^{+3.0}_{-4.2}$ & \cite{LO06.1} \\   
 & & X-ray & $4.0 \pm 0.2$ & & $5.3 \pm 0.3$ & & \cite{XU01.1} \\
Abell 2717 & 0.049 & X-ray & $4.6 \pm 0.3$ & $1.510 \pm 0.089$ & $6.0
\pm 0.3$ & $1.839 \pm 0.122$ & \cite{GA06.1} \\
 & & X-ray & $4.21 \pm 0.25$ & $1.57 \pm 0.19$ & $5.58 \pm 0.32$ &
$1.92 \pm 0.25$ & \cite{PO05.1} \\ 
 & & X-ray & $4.2 \pm 0.3$ & $1.57$ & $5.6 \pm 0.4$
& $1.92$ & \cite{PR05.1} \\
Abell 3562 & 0.0499 & X-ray & $5.4 \pm 0.8$ & & $7.1 \pm 1.0$ & &
\cite{XU01.1} \\ 
Hydra A & 0.0538 & X-ray & $12.3 \pm 0.18$ & $1.02 \pm 0.41$ & $15.9
\pm 0.23$ & $1.15 \pm 0.47$ & \cite{DA01.1} \\
Abell 85 & 0.0557 & X-ray & $4.50$ & $3.36$ & $5.93$ & $4.08$ &
\cite{RI06.1} \\ 
 & & X-ray & $7.5 \pm 0.6$ & & $9.8 \pm 0.8$ & & \cite{XU01.1} \\
Sersic 159 03 & 0.0564 & X-ray & $6.16^{+3.42}_{-2.79}$ &
$2.3^{+7.9}_{-1.4}$ & $8.05^{+4.34}_{-3.56}$ & $2.7^{+10.0}_{-1.7}$
& \cite{VO06.1} \\ 
Abell 2319 & 0.0564 & X-ray & $5.8 \pm 0.2$ & & $7.6 \pm 0.3$ & &
\cite{XU01.1} \\ 
Abell 133 & 0.0569 & X-ray & $4.77 \pm 0.42$ & $4.41 \pm 0.59$ &
$6.28 \pm 0.53$ & $5.33 \pm 0.77$ & \cite{VI05.1} \\
Abell 1991 & 0.0586 & X-ray & $5.78 \pm 0.35$ & $1.63 \pm 0.18$ &
$7.56 \pm 0.45$ & $1.94 \pm 0.22$ & \cite{PO05.1} \\  
 & & X-ray & $5.7^{+0.4}_{-0.3}$ & $1.63$ &
$7.5^{+0.5}_{-0.4}$ & $1.94$ &
\cite{PR05.1} \\   
 & & X-ray & $6.40 \pm 0.46$ & $1.65 \pm 0.24$ & $8.35 \pm 0.58$
 & $1.94 \pm 0.30$ & \cite{VI05.1} \\
Abell 3266 & 0.0594 & X-ray & $3.9 \pm 0.2$ & & $5.2 \pm 0.3$ & &
\cite{XU01.1} \\ 
Abell 3158 & 0.0597 & LOSVD & $2.5^{+0.57}_{-1.8}$ &
$11.4^{+1.7}_{-3.0}$ & $3.3^{+7.2}_{-2.4}$ &
$14.8^{+6.5}_{-5.0}$ & \cite{LO06.1} \\    
Abell 1795 & 0.063 & X-ray & $4.45^{+0.86}_{-0.77}$ &
$7.48^{+2.32}_{-1.58}$ & $5.86^{+1.09}_{-0.98}$ &
$9.07^{+3.03}_{-2.03}$ & \cite{SC06.1} \\ 
 & & X-ray & $4.28^{+2.23}_{-2.41}$ & $8.9^{+54.5}_{-5.6}$ &
$5.64^{+2.84}_{-3.09}$ & $10.8^{+74.4}_{-7.0}$ & \cite{VO06.1} \\
 & & X-ray & $4.82 \pm 0.26$ & $8.38 \pm 0.79$ & $6.32 \pm 0.33$
 & $10.10 \pm 1.01$ & \cite{VI05.1} \\
 & & X-ray & $7.6 \pm 0.3$ & & $9.9 \pm 0.4$ & & \cite{XU01.1} \\
Abell 644 & 0.0704 & X-ray & $4.6 \pm 0.9$ & $7$ & $6.0 \pm 1.2$
& $8$ & \cite{BU05.1} \\
 & & X-ray & $4.6 \pm 0.2$ & & $6.0 \pm 0.3$ & & \cite{XU01.1} \\
Abell 401 & 0.0748 & X-ray & $4.2 \pm 0.3$ & & $5.5 \pm 0.4$ & &
\cite{XU01.1} \\ 
Abell 3112 & 0.0750 & X-ray & $7.06^{+3.62}_{3.23}$ & $2.9^{+13.5}_{-1.9}$
& $9.14^{+2.82}_{-3.05}$ & $3.4^{+16.4}_{-2.2}$ & 
\cite{VO06.1} \\   
Abell 2029 & 0.0767 & X-ray & $6.64^{+0.34}_{-0.38}$ & $7.66^{+0.77}_{-0.58}$ &
$8.60^{+0.42}_{-0.48}$ & $8.97^{+0.94}_{-0.71}$ & \cite{SC06.1} \\
 & & X-ray & $4.38^{+1.64}_{-1.76}$ &
$20.^{+57}_{-16}$ & $5.74^{+2.08}_{-2.24}$ & $24^{+74}_{-20.}$
& \cite{VO06.1} \\  
 & & X-ray & $6.00 \pm 0.30$ & $10.81 \pm 1.08$ & $7.80 \pm
 0.38$ & $12.76 \pm 1.33$ & \cite{VI05.1} \\
 & & X-ray & $4.4 \pm 0.9$ & $12 \pm 2$ & $5.8 \pm 1.1$ & $15 \pm
 3$ & \cite{LE03.1} \\ 
 & & X-ray & $8.4 \pm 0.6$ & & $10.8 \pm 0.8$ & & \cite{XU01.1} \\
RXJ1159.8+5531 & 0.081 & X-ray & $8.3 \pm 2.1$ & $0.787 \pm 0.533$ &
$10.6 \pm 2.6$ & $0.908 \pm 0.686$ & \cite{GA06.1} \\
 & & X-ray & $2.63 \pm 0.43$ & & $3.51 \pm 0.55$ & & \cite{VI05.1} \\
Abell 1651 & 0.0825 & X-ray & $4.9 \pm 0.2$ & & $6.4 \pm 0.3$ & &
\cite{XU01.1} \\ 
Abell 2597 & 0.0852 & X-ray & $5.86 \pm 0.50$ & $3.00 \pm 0.33$ &
$7.59 \pm 0.63$ & $3.54 \pm 0.42$ & \cite{PO05.1} \\    
 & & X-ray & $6.7 \pm 0.6$ & & $8.7 \pm 0.8$ & & \cite{XU01.1} \\
Abell 478 & 0.088 & X-ray & $3.92^{+0.36}_{-0.33}$ &
$13.1^{+2.3}_{-2.1}$ & $5.13^{+0.45}_{-0.41}$ &
$16.0^{+3.0}_{-2.6}$ & \cite{SC06.1} \\ 
 & & X-ray & $2.88^{+2.02}_{- \rightarrow 2.88}$ & $34^{+ \rightarrow
  \infty, a}_{-26}$ & $3.81^{+2.56}_{- \rightarrow 3.81}$ & $43^{+
  \rightarrow \infty, a}_{-33}$ &  \cite{VO06.1} \\
 & & X-ray & $4.22 \pm 0.39$ & $10.8 \pm 1.8$ & $5.52
\pm 0.49$ & $13.1 \pm 2.3$ & \cite{PO05.1} \\
 & & X-ray & $5.33 \pm 0.39$ & $10.53 \pm 1.51$ & $6.92 \pm
 0.49$ & $12.51 \pm 1.88$ & \cite{VI05.1} \\
 & & X-ray & $4.2 \pm 0.4$ & $11$ & $5.5 \pm 0.5$ & $13$ &
 \cite{PO04.1} \\ 
 & & X-ray & $3.67^{+0.31}_{-0.35}$ &
$18.4^{+4.8}_{-2.4}$ & $4.82^{+0.39}_{-0.44}$ &
$22.6^{+6.2}_{-3.1}$ & \cite{AL03.1} \\ 
 & & X-ray & $6.7 \pm 0.4$ & & $8.6 \pm 0.5$ & & \cite{XU01.1} \\
PKS0745$-$191 & 0.103 & X-ray & $5.86^{+1.56}_{-1.07}$ &
 $11.82^{+4.70}_{-3.55}$ & $7.55^{+1.95}_{-1.34}$ &
 $13.89^{+5.85}_{-1.07}$ & \cite{SC06.1} \\ 
 & & X-ray & $5.46^{+3.22}_{-2.88}$ & $9.7^{+52.2}_{-8.5}$ &
$7.05^{+4.04}_{-3.63}$ & $11^{+67}_{-10.}$ & \cite{VO06.1} \\
 & & X-ray & $5.12 \pm 0.40$ & $10.0 \pm 1.2$ & $6.62 \pm 0.50$ &
$11.9 \pm 1.5$ & \cite{PO05.1} \\    
 & & X-ray & $3.83^{+0.52}_{-0.27}$ & $
18.6^{+3.5}_{-4.0}$ & $5.00^{+0.66}_{-0.34}$ & $22.7^{+4.5}_{-5.1}$
& \cite{AL03.1} \\ 
RXJ1416.4+2315 & 0.137 & X-ray & $11.2 \pm 4.5$ & $3.1 \pm 1.0$ &
$14.1 \pm 5.6$ & $3.5 \pm 1.3$ & \cite{KH06.1} \\ 
Abell 1068 & 0.1375 & X-ray & $3.69 \pm 0.26$ & $5.68 \pm 0.49$ &
$4.77 \pm 0.33$ & $6.90 \pm 0.65$ & \cite{PO05.1} \\    
Abell 1413 & 0.143 & X-ray & $4.44^{+0.78}_{-0.75}$ &
$9.31^{+2.69}_{-1.77}$ & $5.69^{+0.97}_{-0.94}$ &
$11.11^{+3.45}_{-2.23}$ & \cite{SC06.1} \\
 & & X-ray & $5.82 \pm 0.50$ & $6.50 \pm 0.65$ & $7.41 \pm 0.62$ & $7.59
\pm 0.82$ & \cite{PO05.1} \\     
 & & X-ray & $4.42 \pm 0.24$ & $10.67 \pm 1.17$ & $5.66 \pm
 0.30$ & $12.73 \pm 1.47$ & \cite{VI05.1} \\
Abell 2204 & 0.152 & WL & $6.3$ & $12^{+3}_{-2}$ & $8.0$ &
 $14^{+3}_{-2}$ & \cite{CL02.1} \\ 
 & & WL & $4.3$ & & $5.5$ & & \cite{CL01.2} \\ 
 & & X-ray & $9.75^{+2.92}_{-2.16}$ & $7.48^{+2.63}_{-1.80}$ &
 $12.2^{+3.60}_{-2.67}$ & $8.44^{+3.14}_{-2.12}$ & \cite{SC06.1} \\
 & & X-ray & $4.59 \pm 0.37$ & $11.8 \pm 1.3$ & $5.86 \pm 0.46$ &
 $14.0 \pm 1.7$ & \cite{PO05.1} \\    
Abell 907 & 0.1603 & X-ray & $5.21 \pm 0.60$ & $6.28 \pm 0.63$ &
$6.61 \pm 0.75$ & $7.37 \pm 0.82$ & \cite{VI05.1} \\
Abell 1689 & 0.18 & SL & $6.0 \pm 0.5$ & $30.$ & $7.6 \pm 0.6$ & $35$
& \cite{HA06.1} \\ 
 & & SL & $5.70^{+0.34}_{-0.50}$ & $130.^{+88}_{-57}$ &
$7.18^{+0.42}_{-0.62}$ & $151^{+104}_{-67}$ & \cite{ZE06.1} \\ 
 & & SL & $6.5^{+1.9}_{-1.6}$ & $34^{+1}_{-2}$ & $8.2^{+2.1}_{-1.8}$
& $40.^{+1}_{-1}$ & \cite{BR05.1} \\ 
 & & WL & $30.4$ & & $37.4$ & & \cite{HA06.1} \\
 & & WL & $22.1^{+2.9}_{-4.7}$ & & $27.2^{+3.5}_{-5.7}$ & & \cite{ME06.1} \\
 & & WL & $3.5^{+0.5}_{-0.3}$ & $14.1^{+6.3}_{-4.7}$ & $4.5^{+0.6}_{-0.4}$
& $17.1^{+7.8}_{-5.8}$ & \cite{BA05.1} \\ 
 & & WL & $11.0^{+1.14}_{-0.90}$ & $17.3 \pm 1.7$ &
 $13.7^{+1.4}_{-1.1}$ & $19.3 \pm 2.0$ & \cite{BR05.2} \\ 
 & & WL & $7.9$ & & $9.9$ & & \cite{CL03.1} \\
 & & WL & $4.8$ & $8.50$ & $6.1$ & $10.0$ &
\cite{KI02.1} \\ 
 & & WL & $6$ & & $8$ & & \cite{CL01.1} \\
 & & WL & $6.0$ & & $7.6$ & & \cite{CL01.2} \\ 
 & & WL+SL & $7.6^{+0.3}_{-0.5}$ & $23$ & $9.5^{+0.4}_{-0.6}$ &
 $26$ & \cite{HA06.1} \\ 
 & & WL+SL & $7.6 \pm 1.6$ & $13.2 \pm 2$ & $9.5 \pm 2.0$ & $15.1
\pm 2$ & \cite{LI06.1} \\
 & & X-ray & $7.7^{+1.7}_{-2.6}$ & & $9.6^{+2.1}_{-3.2}$ & & \cite{AN04.1} \\
Abell 383 & 0.188 & X-ray & $3.76^{+0.53}_{-0.68}$ &
$6.62^{+2.56}_{-1.34}$ & $4.78^{+0.65}_{-0.84}$ &
$7.95^{+3.28}_{-1.68}$ & \cite{SC06.1} \\ 
 & & X-ray & $6.41 \pm 0.57$ & $4.10 \pm 0.47$ & $8.03 \pm 0.70$
 & $4.72 \pm 0.57$ & \cite{VI05.1} \\
MS 0839.9+2938 & 0.194 & X-ray & $6.5 \pm 0.1$ & $6.1$ & $8.1 \pm
0.1$ & $7.0$ & \cite{WA05.1} \\ 
MS 0451.5+0250 & 0.202 & X-ray & $3.79$ & $129$ & $4.80$ & $154$
& \cite{MO99.1} \\ 
Abell 963 & 0.206 & X-ray & $4.39^{+0.88}_{-0.88}$ &
$7.47^{+3.05}_{-1.80}$ & $5.53^{+1.07}_{-1.08}$ &
$8.81^{+3.84}_{-2.21}$ & \cite{SC06.1} \\ 
 & & X-ray & $5.72^{+0.78}_{-1.07}$ &
$7.04^{+1.96}_{-1.26}$ & $7.16^{+0.95}_{-1.31}$ &
$8.14^{+2.43}_{-1.51}$ & \cite{AL03.1} \\ 
RXJ0439.0+0520 & 0.208 & X-ray & $6.66^{+1.53}_{-1.21}$ &
$3.97^{+1.78}_{-1.19}$ & $8.30^{+1.87}_{-1.48}$ &
$4.54^{+2.13}_{-1.40}$ & \cite{SC06.1} \\ 
RXJ1504.1$-$0248 & 0.215 & X-ray & $3.77^{+1.05}_{-1.09}$ &
$17.5^{+13.5}_{-5.6}$ & $4.75^{+1.28}_{-1.34}$ &
$20.9^{+17.3}_{-6.97}$ & \cite{SC06.1} \\ 
MS 0735.6+7421 & 0.216 & X-ray & $6.85$ & $22$ & $8.51$ & $25$
& \cite{MO99.1} \\ 
MS 1006.0+1202 & 0.221 & X-ray & $4.19$ & $31$ & $5.26$ & $36$
& \cite{MO99.1} \\ 
Abell 2390 & 0.230 & X-ray & $2.58 \pm 0.19$ & $16.58 \pm 1.93$ &
$3.28 \pm 0.23$ & $20.45 \pm 2.57$ & \cite{VI05.1} \\
 & & X-ray & $3.20^{+1.79}_{-1.57}$ &
$20.6^{+59.7}_{-11.6}$ & $4.04^{+2.18}_{-1.93}$ &
$24.9^{+79.7}_{-14.4}$ & \cite{AL03.1} \\ 
Abell 2667 & 0.233 & X-ray & $3.02^{+0.74}_{-0.85}$ &
$13.6^{+10.6}_{-4.6}$ & $3.82^{+0.90}_{-1.04}$ &
$16.5^{+13.9}_{-5.8}$ & \cite{AL03.1} \\ 
RXJ2129.6+0005 & 0.235 & X-ray & $4.07^{+2.31}_{-1.97}$ &
$6.46^{+12.6}_{-3.14}$ & $5.09^{+2.80}_{-2.41}$ &
$7.63^{+16.3}_{-3.83}$ & \cite{SC06.1} \\ 
MS 1910.5+6736 & 0.246 & X-ray & $4.65$ & $8.7$ & $5.78$ &
$10.$ & \cite{MO99.1} \\ 
Abell 1835 & 0.252 & WL & $2.96$ & $23.8^{+3.5}_{-3.2}$ & $3.72$ &
$28.8^{+4.2}_{-3.9}$ & \cite{CL02.1} \\ 
 & & WL & $4.8$ & & $5.96$ & & \cite{CL01.2} \\ 
 & & X-ray & $3.42^{+0.45}_{-0.31}$ & $21.2^{+4.62}_{-5.03}$ &
$4.28^{+0.55}_{-0.37}$ & $25.3^{+5.78}_{-6.21}$ & \cite{SC06.1} \\
 & & X-ray & $3.13^{+1.37}_{-1.44}$ & $24^{+104}_{-16}$ &
$3.93^{+1.66}_{-1.76}$ & $29^{+136}_{-20.}$ & \cite{VO06.1} \\ 
 & & X-ray & $4.21^{+0.53}_{-0.61}$ &
$18.2^{+8.4}_{-3.0}$ & $5.24^{+0.64}_{-0.74}$ &
$21.4^{+10.3}_{-3.7}$ & \cite{AL03.1} \\ 
MS 1455.0+2232 & 0.259 & X-ray & $10.9$ & $14$ & $13.2$ & $15$
& \cite{MO99.1} \\ 
Abell 611 & 0.288 & X-ray & $5.08^{+1.72}_{-1.61}$ &
$6.81^{+4.68}_{-2.11}$ & $6.24^{+2.06}_{-1.94}$ &
$7.83^{+5.78}_{-2.53}$ & \cite{SC06.1} \\ 
 & & X-ray & $4.58^{+2.36}_{-2.22}$ &
$9.4^{+16.6}_{-3.9}$ & $5.64^{+2.83}_{-2.68}$ & $11^{+21}_{-5}$ &
\cite{AL03.1} \\ 
Zwicky 3146 & 0.291 & X-ray & $2.32^{+2.31}_{- \rightarrow 2.32}$ &
$28.1^{+ \rightarrow \infty}_{-16.3}$ & $2.91^{+2.78}_{-
  \rightarrow 2.91}$ & $34.5^{+ \rightarrow \infty}_{-20.9}$  &
\cite{SC06.1} \\ 
Abell 2537 & 0.295 & X-ray & $4.83^{+2.32}_{-1.59}$ &
$7.58^{+5.88}_{-3.04}$ & $5.93^{+2.78}_{-1.91}$ &
$8.74^{+7.28}_{-3.64}$ & \cite{SC06.1} \\ 
MS 1008.1$-$1224 & 0.301 & X-ray & $4.40$ & $34$ & $5.40$ &
$39$ & \cite{MO99.1} \\ 
MS 2137.3$-$2353 & 0.313 & SL & $13 \pm 1$ & $2.9 \pm 0.4$ & $16
\pm 1$ & $3.2 \pm 0.4$ & This paper \\
 & & SL & $11.92^{+0.77}_{-0.74}$ &
$7.56^{+0.63}_{-0.54}$ & $14.34^{+0.91}_{-0.88}$ &
$8.29^{+0.71}_{-0.61}$ & \cite{GA05.2} \\  
 & & SL & $12.5^{+5}_{-6}$ & $7.9$ & $15.0^{+6}_{-7}$ & $8.6$ &
 \cite{GA03.1} \\ 
 & & SL & $11.7 \pm 2.1$ & $7.23 \pm 1.90$ & $14.1 \pm 2.5$ &
 $7.93 \pm 2.17$ & \cite{GA02.2} \\
 & & WL & $12^{+12}_{-8}$ & $9.3^{+85.4}_{-7.8}$ &
 $14^{+14}_{-10.}$ & $10.^{+100.}_{-9}$ & \cite{GA03.1} \\ 
 & & WL+SL & $11.73 \pm 0.55$ & $7.72^{+0.47}_{-0.42}$
& $14.11 \pm 0.65$ & $8.47^{+0.53}_{-0.48}$ & \cite{GA05.2} \\  
 & & X-ray & $7.21^{+0.58}_{-0.59}$ & $4.70^{+0.81}_{-0.56}$ &
$8.75^{+0.69}_{-0.71}$ & $5.27^{+0.94}_{-0.65}$ & \cite{SC06.1} \\
 & & X-ray & $5.28^{+2.41}_{-2.52}$ & $8.0^{+32.0}_{-4.8}$ &
$6.44^{+2.87}_{-3.02}$ & $9.1^{+39.0}_{-5.6}$ & \cite{VO06.1} \\
 & & X-ray & $8.71^{+1.22}_{-0.92}$ & $4.25^{+0.84}_{-0.88}$ &
 $10.5^{+1.5}_{-1.1}$ & $4.72^{+0.96}_{-1.00}$ & \cite{AL03.1} \\ 
 & & X-ray & $12.4$ & $11$ & $14.9$ & $12$ & \cite{MO99.1} \\
MACSJ0242.6-2132 & 0.314 & X-ray & $6.69^{+1.23}_{-0.92}$ &
$4.85^{+1.64}_{-1.31}$ & $8.12^{+1.46}_{-1.09}$ &
$5.47^{+1.92}_{-1.51}$ & \cite{SC06.1} \\ 
MS 0353.6$-$3642 & 0.320 & X-ray & $4.84$ & $32$ & $5.91$ &
$36$ & \cite{MO99.1} \\ 
MACSJ2229.8-2756 & 0.324 & X-ray & $7.70^{+3.66}_{-2.62}$ &
$2.74^{+2.02}_{-1.00}$ & $9.30^{+4.34}_{-3.11}$ &
$3.06^{+2.38}_{-1.15}$ & \cite{SC06.1} \\ 
MS 1224.7+2007 & 0.327 & X-ray & $11.3$ & $9.2$ & $13.5$ &
$10.$ & \cite{MO99.1} \\ 
MS 1358.4+6245 & 0.328 & X-ray & $5.84$ & $26$ & $7.09$ & $29$
& \cite{MO99.1} \\ 
ClG 2244$-$02 & 0.33 & SL & $4.3 \pm 0.4$ & $4.5 \pm 0.9$ & $5.2 \pm
0.5$ & $5.2 \pm 1.1$ & This paper \\
MACSJ0947.2+7623 & 0.345 & X-ray & $5.41^{+1.86}_{-1.51}$ &
$10.69^{+8.41}_{-4.04}$ & $6.54^{+2.20}_{-1.79}$ &
$12.15^{+10.04}_{-4.71}$ & \cite{SC06.1} \\ 
MACSJ1931.8-2635 & 0.352 & X-ray & $3.11^{+1.87}_{-1.88}$ &
$16.2^{+ \rightarrow \infty, a}_{-8.6}$ &
$3.81^{+2.22}_{-2.25}$ & $19.2^{+ \rightarrow \infty, a}_{-10.5}$ &
\cite{SC06.1} \\
RXJ1532.9+3021 & 0.3615 & X-ray & $2.77^{+2.28}_{-2.28}$ &
$19^{+675}_{-16}$ & $3.40^{+2.70}_{-2.75}$ & $23^{+1006}_{-19}$ &
\cite{VO06.1} \\  
MACSJ1532.9+3021 & 0.363 & X-ray & $4.71^{+1.32}_{-1.25}$ &
$8.46^{+5.96}_{-2.73}$ & $5.69^{+1.56}_{-1.47}$ &
$9.67^{+7.19}_{-3.22}$ & \cite{SC06.1} \\ 
MS 1512.4+3647 & 0.372 & X-ray & $7.82$ & $7.2$ & $9.35$ &
$7.9$ & \cite{MO99.1} \\ 
MACSJ1720.3+3536 & 0.391 & X-ray & $4.37^{+1.21}_{-0.88}$ &
$9.01^{+4.63}_{-3.30}$ & $5.26^{+1.42}_{-1.04}$ &
$10.31^{+5.55}_{-3.87}$ & \cite{SC06.1} \\ 
ZwCl~0024$+$1652 & 0.395 & WL+SL & $22^{+9}_{-5}$ & $5.7^{+1.1}_{-1.0}$ &
$26^{+10.}_{-6}$ & $6.1^{+1.2}_{-1.1}$ & \cite{KN03.1} \\ 
MACSJ0429.6-0253 & 0.399 & X-ray & $7.64^{+1.57}_{-1.10}$ &
$3.66^{+1.11}_{-0.97}$ & $9.09^{+1.84}_{-1.29}$ &
$4.05^{+1.27}_{-1.10}$ & \cite{SC06.1} \\ 
MACSJ0159.8-0849 & 0.405 & X-ray & $4.93^{+1.01}_{-1.07}$ &
$11.59^{+6.29}_{-3.30}$ & $5.90^{+1.18}_{-1.25}$ &
$13.13^{+7.46}_{-3.84}$ & \cite{SC06.1} \\ 
MS 0302.7+1658 & 0.426 & X-ray & $7.39$ & $8.5$ & $8.75$ &
$9.4$ & \cite{MO99.1} \\ 
MACSJ0329.7-0212 & 0.450 & X-ray & $4.74^{+0.75}_{-0.78}$ &
$6.62^{+2.57}_{-1.56}$ & $5.62^{+0.88}_{-0.91}$ &
$7.48^{+3.03}_{-1.81}$ & \cite{SC06.1} \\ 
RXJ1347.5$-$1145 & 0.451 & WL & $15^{+64}_{-10}$ & $27^{+26}_{-14}$ &
$18^{+74}_{-12}$ & $29^{+31}_{-15}$ & \cite{KL05.1} \\
 & & X-ray & $4.79^{+0.68}_{-0.37}$ &
$32.0^{+6.1}_{-8.2}$ & $5.68^{+0.79}_{-0.43}$ &
$36.1^{+7.1}_{-9.5}$ & \cite{SC06.1} \\ 
 & & X-ray & $4.37^{+1.39}_{-1.24}$ & $33^{+48}_{-18}$ &
$5.20^{+1.62}_{-1.45}$ & $37^{+57}_{-21}$ & \cite{VO06.1} \\ 
 & & X-ray & $6.34^{+1.61}_{-1.35}$ &
$23.7^{+14.2}_{-9.3}$ & $7.49^{+1.87}_{-1.57}$ &
$26.3^{+16.3}_{-10.5}$ & \cite{AL03.1} \\ 
3C 295 & 0.461 & X-ray & $7.79^{+1.04}_{-0.90}$ &
$3.57^{+0.81}_{-0.65}$ & $9.15^{+1.20}_{-0.90}$ &
$3.93^{+0.92}_{-0.73}$  & \cite{SC06.1} \\ 
 & & X-ray & $7.90^{+1.71}_{-1.72}$ &
$37.6^{+15.9}_{-10.2}$ & $9.28^{+1.98}_{-1.99}$ &
$41.3^{+18.1}_{-11.4}$ & \cite{AL03.1} \\  
MACSJ1621.6+3810 & 0.461 & X-ray & $5.97^{+2.95}_{-1.94}$ &
$7.10^{+5.33}_{-2.90}$ & $7.05^{+3.42}_{-2.26}$ &
$7.91^{+6.25}_{-3.31}$ & \cite{SC06.1} \\ 
MACSJ1311.0-0311 & 0.494 & X-ray & $4.42^{+1.39}_{-1.05}$ &
$6.22^{+3.71}_{-2.15}$ & $5.22^{+1.60}_{-1.22}$ &
$7.02^{+4.38}_{-2.49}$  & \cite{SC06.1} \\ 
MACSJ1423.8+2404 & 0.539 & X-ray & $7.69^{+0.70}_{-0.79}$ &
$5.28^{+1.13}_{-0.76}$ & $8.92^{+0.81}_{-0.91}$ &
$5.77^{+1.27}_{-0.84}$ & \cite{SC06.1} \\ 
MS 0015.9+1609 & 0.546 & X-ray & $4.37$ & $93.3$ & $5.11$ &
$105$ & \cite{MO99.1} \\  
MS 0451.6$-$0305 & 0.55 & SL & $5.5 \pm 0.3$ & $18 \pm 2$ & $6.4 \pm
0.3$ & $20. \pm 2$ & This paper \\
3C 220.1 & 0.62 & SL & $4.3 \pm 0.2$ & $3.1 \pm 0.3$ & $5.0 \pm 0.2$ &
$3.5 \pm 0.3$ & This paper \\
SDSS J1004+4112 & 0.68 & SL & $5$ & $3.87$ & $6$ & $4.25$ &
\cite{WI04.1} \\ 
MACSJ0744.9+3927 & 0.686 & X-ray & $4.32^{+1.43}_{-1.06}$ &
$8.83^{+4.84}_{-3.16}$ & $4.95^{+1.61}_{-1.20}$ &
$9.78^{+5.60}_{-3.58}$ & \cite{SC06.1} \\ 
MS~1137.5+6625 & 0.783 & SL & $3.3 \pm 0.2$ & $6.5 \pm 0.7$ & $3.8 \pm 0.2$ &
$7.2 \pm 0.8$ & This paper \\
ClJ 1226.9+3332 & 0.89 & X-ray & $7.9^{+1.7}_{-1.4}$ &
$6.8^{+1.6}_{-1.2}$ & $8.8^{+1.9}_{-1.5}$ & $7.2^{+1.7}_{-1.3}$ &
\cite{MA06.1} \\
\label{tbl-1}
\end{longtable}
\end{center}

\end{appendix}

\end{document}